\documentclass[pra,aps,epsfig,showpacs,twocolumn,floats,superscriptaddress]{revtex4}
\usepackage{mathrsfs}
\usepackage{amssymb,color}
\usepackage{amssymb,graphicx,amsmath}
\begin{document}

\title{The dynamics of a qubit coupled strongly with a quantum oscillator
}
\author{Guihua Tian}
\email{tgh-2000@263.net, tgh20080827@gmail.com, }
 \affiliation{School of Science, Beijing
University of Posts And Telecommunications.
\\ Beijing 100876 China.}

\begin{abstract}
The dynamics of a qubit coupled with a quantum oscillator is
re-studied in the region of strong coupling. The non-degenerate
perturbation is added to the usual degenerate one and new results
are given.
\end{abstract}

\pacs{ 42.50.Md, 42.50.Hz, 85.25.Cp} \maketitle

\section{Introduction}
A qubit coupled with a quantum oscillator is the simplest model in
quantum system, yet is an abundance of  applications both in
theoretical and practical aspects. Its dynamics is described by the
Rabi Hamiltonian \cite{rabi}. However, its analytical solutions is
still unavailable. Many approximations are applied. Among them, the
rotation wave approximation (RWA) \cite{mand} and adiabatic
approximation (AA) are major ones. Under weak coupling and near
resonant condition, it has been solved analytically by the method of
RWA by discarding the non-energy-conserving terms, and is also
called the Jaynes-Cumming model in quantum optics\cite{mand}. In the
opposite condition, that is, the strong coupling and large detuning,
it is investigate by the adiabatic approximation (AA) and
generalized RWA \cite{iris}-\cite{cres}. The AA method assumed large
detuning and first only treat the quantum oscillator influenced by
the qubit through the coupling, then solve the whole system by
treating the qubit as a perturbation. Further assumption of the
qubit  must be made for its solutions in AA method, that is why
sometimes called it adiabatic approximation. The adiabatic
assumption requires that the qubit as a perturbation does not have
influence on the transition of the quantum oscillator from its
different number states and discards these transition terms
concerning the displaced number states and solved it accordingly.

Because recent achievements in circuit QED make  the strong or even
the ultrastrong coupling regime of light matter interaction
achievable\cite{iris}-\cite{albe2},  this in turn renews the
theoretical interesting in qubit under adiabatic approximation.
Further extensions are made by replacing the qubit by two qubit and
N-qubits or linear coupling by non-linear ones, etc
\cite{agar}-\cite{albe2}.

In the paper, we will restudy the qubit coupled strongly with
quantum oscillator and focus on the rationality of  the adiabatic
assumption. Though physically the low energy of the qubit can not
promotes the transition of the oscillator to its different displaced
number states, these omitting terms  sometimes are larger than the
keeped terms of the transition of the same number states displaced
differently. So we will make a scrutiny into the adiabatic
approximation and a generalized approximation is given and its new
results are discussed. Section 2 reviews the adiabatic approximation
in the study of the qubit and quantum oscillator coupled system and
section 3 gives new approximation. Final section will discuss and
conclude the results.

\section{The brief introduction of the adiabatic approximation }

The Rabi Hamiltonian for  a qubit interacting with a  harmonic
oscillator is \cite{iris}, \cite{agar},
\begin{eqnarray}\label{H for qutrit}
     H=\frac12 \hbar\omega_{0}\sigma_z+ \hbar\omega a^{\dagger}a
+ \hbar\beta(a+a^{\dagger})\sigma_x , \end{eqnarray} where
$\sigma_x,\ \sigma_z$ are the usual Pauli matrices with
$\sigma_x=\sigma_++\sigma_-$.

In the paper, we do not consider the RWA case where the
non-rotation-terms $a^+\sigma_+ ,\ a\sigma_-$ are neglected. Under
large detuning and strong coupling condition, the qubit is regarded
as a perturbation. The so-called free hamiltonian is
\begin{eqnarray}\label{H for qutrit}
     H_0= \hbar\omega a^{\dagger}a
+ \hbar\beta(a+a^{\dagger})\sigma_x,  \end{eqnarray}whose solutions
are \begin{eqnarray}
      H_0|N_m,m\rangle &=& (N-\beta^2)\hbar \omega|N_m,m\rangle, \\
      |N_m,m\rangle &=& |m\rangle|N_m\rangle ,\  m=+,\ -,
    \end{eqnarray}
where $|+\rangle,\ |-\rangle$ are the eigenvectors of the operator
$\sigma_x$ with $\sigma_x|+\rangle=|+\rangle,\ \sigma_x|-\rangle=-
|-\rangle$, and $|N_m\rangle$ are the displaced number states for
the quantum oscillator, that is
\begin{eqnarray}
|N_{\pm}\rangle=\hat{D}(\mp \beta)|N\rangle,\ \
\hat{D}(\beta)=e^{\beta(a^{\dagger}-a)}.
\end{eqnarray}
$|+\rangle,\ |-\rangle$  are not $|\uparrow\rangle,\
|\downarrow\rangle$ , which are the eigenvectors of the operator
$\sigma_z$.

The Hamiltonian describes the quantum oscillator influenced by force
from its coupling with with qubit, and its solutions show that the
force of the qubit acting on the oscillator has make the oscillator
displace its equilibrium position according to the states of the
qubit, just as shown in the displaced number states
$|\psi_{n,m}\rangle$ in its solutions.

The displaced number states $|N_{\pm}\rangle$ are not completely
orthogonal due to the fact
\begin{eqnarray}
\langle N_{\mp}|N_{\pm}\rangle\ne 0 \ with \ m\ne m'.
\end{eqnarray}
This will result in the complex situation when including the
perturbation term $\frac12\hbar \omega_0 \sigma_z$ of the qubit to
be treated. This term $\frac12\hbar \omega_0 \sigma_z$ of the qubit
will produces the transition of the oscillator from its various
eigenstates $|N_{\pm}\rangle$. From the theoretical view, the large
detuning ($\omega_0\ll \omega$) will make it reasonable to omit its
transition among different numbers states. This means that the qubit
mainly results in transition of the oscillator from the same number
states  which are displaced differently induced by the corresponding
states of the qubit , that is, the transitions of the form
$|N_{\pm}\rangle\rightarrow |N_{\mp}\rangle$. The other transitions
($|N_{\pm}\rangle\rightarrow |N'_{\mp}\rangle ,N\ne N',$) will be
omitted from the consideration. This approximation is called
adiabatic. In this way, the whole Hamiltonian  becomes
block-diagonal as
\begin{eqnarray}
H=\hbar\omega \left(
                \begin{array}{cccccc}
                  H_0 & 0 &\cdots & 0 & 0 & \cdots \\
                  0 & H_1 &\cdots & 0 & 0 & \cdots \\
                  \cdots & \cdots & \cdots & \cdots & \cdots & \cdots \\
                  0 & 0 & \cdots & H_N & 0  & \cdots \\
                  \cdots & \cdots & \cdots & \cdots & \cdots & \cdots \\
                \end{array}
              \right)
\end{eqnarray}
with the $2\times 2$ matrices $H_N$ defined as
\begin{eqnarray}
H_N=\left(
      \begin{array}{cc}
        N -\beta^2&  \Omega_{N} \\
        \Omega_{N} & N-\beta^2 \\
      \end{array}
    \right),
\end{eqnarray}
where
\begin{eqnarray}
  \Omega_{N} &=& \frac{\omega_0}{2\omega}\langle +|\sigma_z|-\rangle\langle N_{+}|N_-\rangle\nonumber \\
  &=&  \frac{\omega_0}{2\omega}\exp{(-2\beta^2)}L_N(4\beta^2).
  \end{eqnarray}
$L_N(x)$ in the above equation are the Laguerre polynomials. The new
solutions are easy to obtain:
\begin{eqnarray}
 E^0_{N,\pm} &=& \hbar\omega\left(N-\beta^2\pm \Omega_{N}\right), \label{old en pm}\\
 |E^0_{N,\pm}\rangle &=& \frac1{\sqrt2}\left(|N_{+},+\rangle\pm
 |N_{-},-\rangle\right).\label{old en vs}
                    \end{eqnarray}
All eigen-values $E^0_{N,\pm} $ are influenced by the parameters
$\beta,\ \frac{\omega_0}{\omega},\ N$, so are the eigenvectors
$|E^0_{N,\pm}\rangle $.

The eigenvectors $|E^0_{n,\pm}\rangle $ are complete and orthogonal
basis for the composite system of the qubit and the quantum
oscillator. The evolution of the qubit will definitely depends on
the initial states of the composite system. Here we only give some
examples for showing the dynamics of the qubit. When the initial
states of the composite system are $E^0(0)=|\Psi_{N,+}\rangle $, the
possibility of the qubit remaining in its initial state are
\begin{eqnarray}
P(t)=\cos^2(\Omega_N*\omega t).
\end{eqnarray}
\begin{figure}
\includegraphics[width=0.4\textwidth]{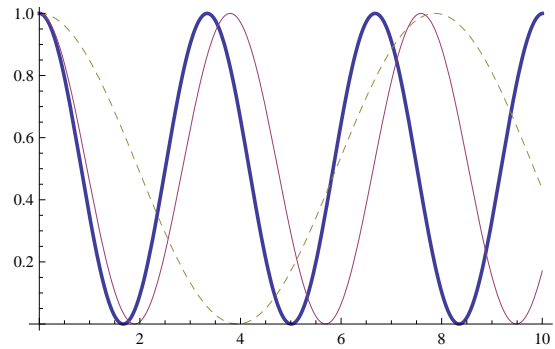}
\caption{The dynamics of the system depends on the parameters
$n,\protect\beta $. The thick, solid and dashed lines correspond to
the figures of $P(t)$ with $N=1,\ \beta=0.2$, $N=4,\ \beta=0.2$,
$N=1,\ \beta=0.7$ respectively.}\label{fig1}
\end{figure}
The general initial state will evolute by the principle of
superposition, see details in Ref.\cite{iris}.

\section{The perturbation restudied in details}

\begin{figure}
\includegraphics[width=0.4\textwidth]{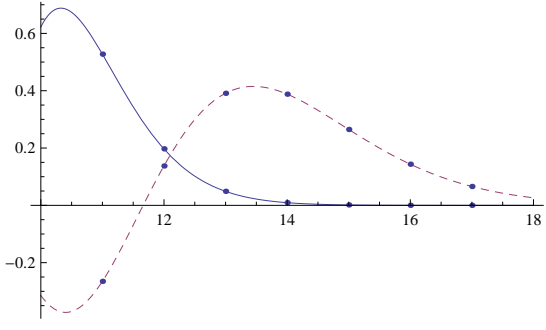}
\caption{The comparison of the various perturbation terms $\langle
10_{-}|\frac12 \hbar\omega_0\protect\sigma_z|N'_{+}\rangle$. The
solid  and dashed lines correspond to the figures for $\langle
10_{-}|\frac12 \hbar\omega_0\protect\sigma_z|N'_{+}\rangle$ with
$\protect\beta=0.2,\ 0.7$ respectively. The independent variable is
$N'$.}\label{fig2}
\end{figure}

As stated before, the adiabatic approximation omits various terms of
the form $$|N_{\pm}\rangle\rightarrow |N'_{\mp}\rangle , \ N\ne N'$$
under the physical ground that $\omega_0\ll \omega$. Though
physically the qubit seems not able to promote the oscillator to go
from one displaced number state to a new one with different number
as its energy gap $|n-n'|\hbar\omega_0$ is far more less that
$\hbar\omega$ of the transition $|N_{+}\rangle\rightarrow
|N'_{+}\rangle,  \ N\ne N'$ or $|N_{-}\rangle\rightarrow
|N'_{-}\rangle,  \ N\ne N'$ , however, this is not true for the
transition of the type $|N_{+}\rangle\rightarrow |N'_{-}\rangle,  \
n\ne n'$ or $|N_{-}\rangle\rightarrow |N'_{+}\rangle,  \ N\ne N'$ .
Detailed quantities study also support the conclusion, and shows
that the perturbation terms $\langle
N_{\pm}|\frac12\hbar\omega_0\sigma_z|N'_{\pm}\rangle$ depend
non-linearly on the coupling strength $\beta$ and the parameters
$N,N'$. Given an arbitrarily strong coupling $\beta$, it is
generally true that $\langle N_{\mp}|\frac12
\hbar\omega_0\sigma_z|N_{\pm}\rangle$ will be larger than most of
 $\langle N'_{\mp}|\frac12
\hbar\omega_0\sigma_z|N_{\pm}\rangle,\ N\ne N'$ . However, there are
still several numbers $N'=N_i,\ i=1,2,\cdots, N_k$ that  $\langle
N'_{\mp}|\frac12 \hbar\omega_0\sigma_z|N_{\pm}\rangle$  are much
larger than   $\langle N_{\mp}|\frac12
\hbar\omega_0\sigma_z|N_{\pm}\rangle$ . Quantity study shows the
number $N_k$ is finite and depends both the coupling strength
$\beta$ and the number $N$, so it will change accordingly. For
example, Fig.(\ref{fig2}) shows   $\langle 13_{+},+|\frac12
\hbar\omega_0\sigma_z|10_{-},- \rangle$  is greater than $\langle
10_{+},+|\frac12 \hbar\omega_0\sigma_z|10_{-},- \rangle$ when
$\beta=0.7$. So the adiabatic approximation must add these terms to
be more applicable.

There are two kinds of perturbations in quantum theory. The
perturbation to the degenerate system and that to the non-degenerate
ones. Let's reconsider the perturbation  $\hbar\omega_0\sigma_z$ to
the free Hamiltonian $H_0$. The eigenstates are $|N_+,+\rangle , \
|N_-,-\rangle , \ N=0,1,2,\cdots $. The states  with the same
numbers $n$ are degenerate, while other states with different
numbers are not. So the perturbation $\frac12\hbar\omega_0\sigma_z$
will relate the two kinds ones in quantum theory. Obviously, the
adiabatic perturbation is the one only applied to the  degenerate
cases of the free Hamiltonian.      So, it is not complete as there
is the non-degenerate perturbation to need to be added.

Some observation is given here for convenience: whenever there are
perturbations of combination of both the degenerate one and
non-degenerate one, one must first treat the degenerate one, then
treats the non-degenerate one. The reverse order can not work, as
the condition of perturbation is small relative the original energy
gap of the free Hamiltonian can not be met due to its degeneration.
Further more, for the second perturbation, there are two bases to
use: one is from the eigen-states of the the free hamiltonian, the
other comes from the first degenerate perturbation. In our case,
they are $|N_+,+\rangle , \ |N_-,-\rangle , \ N=0,1,2,\cdots, $ and
$|E^0_{N,\pm}\rangle  , \ N=0,1,2,\cdots, \ i=+,-$. Of course, the
basis $|E^0_{N,\pm}\rangle  , \ N=0,1,2,\cdots,\ \ i=+,- $ are more
favorable for the calculation of the second perturbation.

By the use of the basis $|E^0_{N,\pm}\rangle  , \ N=0,1,2,\cdots $,
the dynamic equation for the composite systems is \begin{eqnarray}
H|E_{N,+}\rangle=E_{N,+}|E_{N,+}\rangle,\
H|E_{N,-}\rangle=E_{N,-}|E_{N,-}\rangle.\label{hn eq}\end{eqnarray}
Then the corresponding eigen-vectors and eigen-energies are

\begin{eqnarray}
|E_{N,+}\rangle
&=&|E^0_{N,+}\rangle+\sum_{I=1}^{N_k}(a^+_{N,I}|E^0_{I,+}\rangle
+b^+_{N,I}|E^0_{I,-}\rangle),\\ |E_{N,-}\rangle
&=&|E^0_{N,-}\rangle+\sum_{I=1}^{N_k}(a^-_{N,I}|E^0_{I,+}\rangle+b^-_{N,I}|E^0_{I,-}\rangle),
\label{new en vs}\\
E_{N,+}&=&\hbar\omega\bigg[(N-\beta^2+\Omega_N)
+\sum_{I=1}^{N_k}\frac{|\langle N_+|
I_-\rangle|^2}{4(N-I)}\frac{\omega^2_0} {\omega^2}\bigg],\label{new en plus}\\
  E_{N,-}&=&\hbar\omega\bigg[(N-\beta^2-\Omega_N)
+\sum_{I=1}^{N_k}\frac{|\langle N_+|
I_-\rangle|^2}{4(N-I)}\frac{\omega^2_0} {\omega^2}\bigg],\label{new
en mins}
\end{eqnarray}
where
\begin{eqnarray}
  a^+_{N,I} &=& - b^-_{N,I}=\frac{(1+(-1)^{N-I})\langle I_+| N_-\rangle}{4(N-I)}\frac{\omega_0}{\omega},\\
  b^+_{N,I}&=&-a^-_{N,I}=\frac{(1-(-1)^{N-I})\langle I_+|
  N_-\rangle}{4(N-I)}\frac{\omega_0}{\omega}.\label{a bni}
  \end{eqnarray}
  The summation includes those terms of $\langle N'_{\mp}|\frac12 \hbar\omega_0\sigma_z|N_{\pm}\rangle$ that are
  comparable with the term $\Omega_N=\langle N_{\mp}|\frac12 \hbar\omega_0\sigma_z|N_{\pm}\rangle $. See Appendix for details.
\section{Discussion and Conclusion}
\begin{figure}
\includegraphics[width=0.4\textwidth]{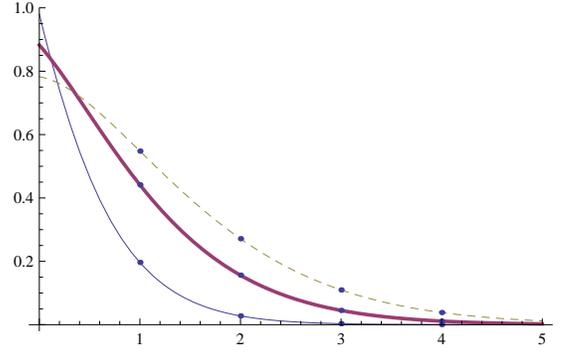}
\caption{The comparison of the various perturbation terms $\langle
0_{-}|N'_{+}\rangle$. The solid , thick and dashed lines of the
figures represent for $\langle0_{-} |N'_{+}\rangle$ with
$\protect\beta=0.2,\ 0.5,\ 0.7$ respectively. The independent
variable is $N'$.}\label{fig3}
\end{figure}
From comparison of the results (\ref{new en vs}),(\ref{new en
plus})-(\ref{new en mins}) with the old ones (\ref{old en vs}),
(\ref{old en pm}), we could see the justification of the adiabatic
approximation. The parameter $\frac{\omega_0}{\omega}$ is small, the
new eigenstates (\ref{new en vs})  differ from the old one (\ref{old
en vs}) only on first order of $\frac{\omega_0}{\omega}$. Though the
new eigenvalues only has some extra terms in the second order of
$\frac{\omega_0}{\omega}$, the old one (\ref{old en pm}) is already
up to the first order of $\frac{\omega_0}{\omega}$. Some terms
$\langle N'_{\mp}|\frac12 \hbar\omega_0\sigma_z|N_{\pm}\rangle$ may
be comparable or even  larger  than $\langle N_{\mp}|\frac12
\hbar\omega_0\sigma_z|N_{\pm}\rangle$, discarding these terms to
make the whole Hamiltonian block-diagonal generally is feasible
under the large detuning condition. However, because the term
$\Omega_n$ depends non-linearly on the parameter $\beta$, in case of
it being zero by some chosen $\beta$, that is at the critical
points, we  consider it is reasonable to refer the adiabatic
approximation as  the new results (\ref{new en vs}),(\ref{new en
plus})-(\ref{new en mins}).

Here we give an example. Consider the composite system is initially
in state $\Psi(0)=|+\rangle |0_+\rangle=|0_+,+\rangle$. The quantum
oscillator is in its displaced vacuum state or coherent state
$|0_+\rangle=\hat{D}(-\beta)|0\rangle$. Suppose the coupling
strength $\beta=0.2$, we will have
\begin{eqnarray}
|\Psi(t)\rangle=\frac1{\sqrt2}[e^{-i\frac{E^0_{0,+}}{\hbar}t}
|0_+,+\rangle+ e^{-i\frac{E^0_{0,-}}{\hbar}t} |0_-,-\rangle ],\\
E^0_{0,+}=\hbar\omega(-\beta^2+\Omega_0),\
E^0_{0,-}=\hbar\omega(-\beta^2-\Omega_0),\\
\Omega_0=\frac{\omega_0}{2\omega}\langle 0_-|0_+\rangle\label{old
sol}
\end{eqnarray}
as the state of the system at time $t$ by the old adiabatic
approximation method. From Fig.(\ref{fig3}), we obtain that $\langle
0_-|0_+\rangle=0.98099,\ \langle 0_-|1_+\rangle=0.19604,\ \langle
0_-|2_+\rangle=0.0277242,\ \langle 0_-|3_+\rangle=0.00320132 $ and
all others much smaller than $0.00320132$. So, in the new results of
 Eqs.(\ref{new en vs}), we have
\begin{eqnarray}|E_{0,+}\rangle
=|E^0_{0,+}\rangle+(a^+_{0,2}|E^0_{2,+}\rangle
+b^+_{0,1}|E^0_{1,-}\rangle)
+b^+_{0,3}|E^0_{3,-}\rangle).\end{eqnarray} If we further choose
$\frac{\omega_0}{\omega}=0.3$, then
\begin{eqnarray}|E_{0,+}\rangle
& \approx
&|E^0_{0,+}\rangle+0.0029406|E^0_{1,-}\rangle.\end{eqnarray}
Similarly, we have \begin{eqnarray}|E_{0,-}\rangle & \approx
&|E^0_{0,-}\rangle-0.0029406|E^0_{1,+}\rangle.\end{eqnarray} The
interesting things   of the above equations are that
$|E_{0,+}\rangle$ combines the two states $|E^0_{0,+}\rangle,\
|E^0_{1,-}\rangle$ and $|E_{0,-}\rangle$ combines the two states
$|E^0_{0,-}\rangle,\ |E^0_{1,+}\rangle$ with opposite coefficient.
Of course, the coefficient in front of $|E^0_{1,-}\rangle$ or
$|E^0_{1,+}\rangle$  very small. The change of the eigenvalues is
ignorable in this case, that is, $E_{N,\pm}\approx E^0_{N,\pm}$.
Therefore, $\Psi(t)=e^{-i\frac{\hat{H}}{\hbar}t}|0_+,+\rangle$
becomes
\begin{eqnarray}\Psi(t)&\approx &
\frac1{\sqrt2}\left(e^{-i\frac{E^0_{0,+}}{\hbar}t}
|E^0_{0,+}\rangle+ e^{-i\frac{E^0_{0,-}}{\hbar}t}
|E^0_{0,-}\rangle\right)\nonumber\\
&+&\frac{0.0029406}{\sqrt2}\left[(e^{-i\frac{E^0_{0,+}}{\hbar}t}-e^{-i\frac{E^0_{1,-}}{\hbar}t})
|E^0_{1,-}\rangle\right]\\ &-&
\frac{0.0029406}{\sqrt2}\left[(e^{-i\frac{E^0_{0,-}}{\hbar}t}-e^{-i\frac{E^0_{1,+}}{\hbar}t})|E^0_{1,+}\rangle\right]\label{new
sol}.\end{eqnarray} Eq.(\ref{new sol}) has the extra term, its last
two  parts, compared tith the old one (\ref{old sol}). So, even the
quantum oscillator is initially in its displaced vacuum states, it
will have some possibility to evolve into its displaced one photon
states. The result may be of some use for the application of qubits
coupled with a quantum oscillator.

In summary, concerning the system of a qubit coupled strongly to a
quantum oscillator, the adiabatic approximation method is restudied
and is extended or modified to the second order of
$\frac{\omega_0}{\omega}$ for the eigenvalues and the first order
for the eigenstates. Though the modification is small and the old
adiabatic approximation is justified, the new results will have some
application. We could also extend the study to the system of
$N-$qubits coupled with a quantum oscillator, where the spectrum of
the free Hamiltonian $H_0$ is $E^0_{N,m}=\hbar\omega(1-m^2\beta^2),\
m=0,\ \pm1,\cdots,\ N=0,1,\cdots $. The smallest energy gap is
$\Delta E^0=\hbar\omega(1-\Delta m^2\beta^2)$. For example, $\Delta
m^2=m_2^2-m_1^2=3$, the condition that the qubits energy
$\hbar\omega_0$ is much more smaller than the transition energy gap
$\Delta E=\hbar\omega (1-3\beta^2)$ can not be met for strong
coupling (large $\beta$). So it is plausible  the inclusion of the
$N$-qubits will not induce the the transition of different displaced
number states, that is, the adiabatic approximation might be put
into doubt in this case. Our study in the paper must be used to see
that the adiabatic approximation can still be justified. This will
be investigated in ref.\cite{tian}.
 \acknowledgments  The work was supported
by the National Natural Science of China (No. 10875018).

\section{Appendix}
By the use of the perturbation method in quantum mechanics, we could
solve Eq.(\ref{hn eq}) by the following
\begin{widetext}
\begin{eqnarray}
|E_{N,+}\rangle
=|E^0_{N,+}\rangle+\sum_{I=1}^{N_k}a^+_{N,I}|E^0_{I,+}\rangle
+\sum_{I=1}^{N_k'}b^+_{N,I}|E^0_{I,-}\rangle,\ \ |E_{N,-}\rangle
=|E^0_{N,-}\rangle+\sum_{I=1}^{N_j}a^-_{N,I}|E^0_{I,+}\rangle+\sum_{i=1}^{n_j'}b^-_{N,I}|E^0_{I,-}\rangle
\label{new en vs1},
\end{eqnarray}
where
\begin{eqnarray}
  a^+_{N,I} &=& \frac{\omega_0}{2(N-I)\omega}\langle E^0_{I,+}|\hat{\sigma}_z|E^0_{N,+}\rangle=\frac{\omega_0}{4(N-I)\omega}(\langle I_+| N_-\rangle+\langle
  I_-| N_+\rangle),\\
  b^+_{N,I}&=&\frac{\langle E^0_{I,-}|\hat{\sigma}_z|E^0_{N,+}\rangle}{(N-I)\hbar\omega}=\frac{\omega_0}{4(N-I)\omega}(\langle I_+| N_-\rangle-\langle
  I_-| N_+\rangle),
   \\
  a^-_{N,I} &=& \frac{\langle E^0_{I,+}|\hat{\sigma}_z|E^0_{N,+}\rangle}{(N-I)\hbar\omega}=\frac{\omega_0}{4(N-I)\omega}(-\langle I_+| N_-\rangle+\langle
  I_-| N_+\rangle),\\
  b^-_{N,I}&=&\frac{\langle E^0_{I,+}|\hat{\sigma}_z|E^0_{N,+}\rangle}{(N-I)\hbar\omega}=\frac{\omega_0}{4(N-I)\omega}(-\langle I_+| N_-\rangle-\langle
  I_-| N_+\rangle),
\end{eqnarray}
and
\begin{eqnarray}
E_{N,+}&=&E^0_{N,+}+\frac14
\hbar\omega_0\sum_{I=1}^{N_k}a^+_{N,I}(\langle N_+|
I_-\rangle+\langle
  N_-| I_+\rangle)
+\frac14 \hbar\omega_0\sum_{I=1}^{N_k'}b^+_{N,I}(-\langle N_+|
I_-\rangle+\langle
 N_-| I_+\rangle)\nonumber\\
 &=&\hbar\omega\bigg[(n-\beta^2+\Omega_n)  +\sum_{I=1}^{N_k}\frac{\omega^2_0}
{(N-I)16\omega^2}|\langle N_+| I_-\rangle+\langle
  N_-| I_+\rangle|^2+\sum_{I=1}^{N'_k}\frac{\omega^2_0}
{(N-I)\omega^2}|\langle N_+| I_-\rangle-\langle
  N_-| I_+\rangle|^2\bigg]\label{new en plus1},\\
  E_{N,-}&=&E^0_{N,-}+\frac14
\hbar\omega_0\sum_{I=1}^{N_j}a^-_{N,I}(\langle N_+|
I_-\rangle-\langle
  N_-| I_+\rangle)
 +\frac14 \hbar\omega_0\sum_{I=1}^{N'_j}b^-_{N,I}(-\langle N_+|
I_-\rangle-\langle
 N_-| I_+\rangle)\nonumber\\
&=&\hbar\omega\bigg[(N-\beta^2-\Omega_N)
+\sum_{I=1}^{N_k}\frac{\omega^2_0} {(N-I)16\omega^2}|\langle N_+|
I_-\rangle-\langle
  N_-| I_+\rangle|^2+\sum_{I=1}^{N'_k}\frac{\omega^2_0}
{16(N-I)\omega^2}|\langle N_+| I_-\rangle+\langle
  N_-| I_+\rangle|^2\bigg].\label{new en mins1}
\end{eqnarray}
\end{widetext}
In all the above equations, the summations all mean the sum of the
terms involved the quantities $\langle N'_{\mp}|\frac12
\hbar\omega_0\sigma_z|N_{\pm}\rangle$ that can not be ignorable.
Usually, there are just several such terms need to calculate.

Because of the fact \cite{iris} \begin{eqnarray}\langle M_+|
N_-\rangle=e^{-2\beta^2}(-\beta)^{m-n}\sqrt{\frac{M!}{N!}}L_N^{M-N}(4\beta^2),\
M\ge N,\\
\langle M_+|
N_-\rangle=e^{-2\beta^2}(-\beta)^{N-M}\sqrt{\frac{M!}{N!}}L_M^{N-M}(4\beta^2),\
M< N,\end{eqnarray} and $\langle M_+| N_-\rangle$ being real, there
are the relations $$\langle M_+| N_-\rangle=(-1)^{N-M}\langle M_-|
N_+\rangle$$ and $$\langle M_-| N_+\rangle=(-1)^{N-M}\langle N_-|
M_+\rangle,$$ the above results concerning the coefficients
$a^+_{N,I}$, etc, could be simplified as Eqs.(\ref{a bni}). Please
note that  $L_N^{K}(x)$ in above is the associated Laguerre
function.

The simple relation among the coefficients $a^+_{N,I}, a^-_{N,I} \
b^+_{N,I} ,\ b^-_{N,I}  $ shows that the all upper limits of
summation in Eqs (\ref{new en vs})-(\ref{new en mins}) could be put
to the same as $N_k$ which depends on both $N,\ I$. So, the results
could be concisely written as Eqs (\ref{new en vs})-(\ref{new en
mins}).

\acknowledgments  The work was partly supported by the National
Natural Science of China (No. 10875018) and the Major State Basic
Research Development Program of China (973 Program:
No.2010CB923200).

\end{document}